\documentclass[11pt,titlepage]{article}

\usepackage{setspace} 
\usepackage{color}
\usepackage{ulem}

\usepackage{amsmath}

\usepackage{graphicx}

\usepackage[round,numbers,sort&compress]{natbib} 
\title{ Mapping the energy and diffusion landscapes of membrane proteins at the cell surface using high-density single-molecule imaging and Bayesian inference: application to the multi-scale dynamics of glycine receptors in the neuronal membrane.}

\author{Jean-Baptiste~Masson\thanks{
           Corresponding author. email: jbmasson@pasteur.fr. Address: 
           Institut Pasteur,
            Physics of Biological Systems, 
            28 rue du Dr Roux, CNRS, UMR 3525
            75724, Paris Cedex 15,
             France. 
         Tel.:~+33 (0)1 40 61 39 23. 
         $^{*}$ These authors contributed equally } \\
	Physics of Biological System, \\
	Pasteur Institute, Paris , France,\\
	CNRS UMR 3525, Paris, France.\\ 
	\and Patrice Dionne* \\
	Laboratoire Kastler Brossel, CNRS UMR 8552,\\
	  Ecole Normale Superieure, Paris , France,\\
	 Centre de recherche Université Laval Robert-Giffard, Quebec, Canada.\\
	\and Charlotte Salvatico* \\
	Biologie Cellulaire de la Synapse,\\
	Institut National de la Sante et de la Recherche Medicale U1024, \\ 
         Institut de Biologie de l'Ecole Normale Superieure (IBENS),\\
          75005 Paris, France. \\
	\and Marianne Renner \\
	Biologie Cellulaire de la Synapse,\\
	Institut National de la Sante et de la Recherche Medicale U1024, \\ 
         Institut de Biologie de l'Ecole Normale Superieure (IBENS),\\
          75005 Paris, France. \\
	\and Christian G. Specht \\
	Biologie Cellulaire de la Synapse,\\
	Institut National de la Sante et de la Recherche Medicale U1024, \\ 
         Institut de Biologie de l'Ecole Normale Superieure (IBENS),\\
          75005 Paris, France. \\
	\and Antoine Triller \thanks{Corresponding author, email: triller@biologie.ens.fr, Address: 
           Ecole Normale Superieure, 
           Biologie Cellulaire de la Synapse
           Institut National de la Sante et de la Recherche Medicale U1024
           Institut de Biologie de l'Ecole Normale Superieure (IBENS),
           45 rue d'Ulm,
           Tel.:~ +33 (0)1 44 32 35 47.} \\
	Biologie Cellulaire de la Synapse,\\
	Institut National de la Sante et de la Recherche Medicale U1024, \\ 
         Institut de Biologie de l'Ecole Normale Superieure (IBENS),\\
          75005 Paris, France. \\
	\and Maxime Dahan \thanks{Corresponding author, email: maxime.dahan@curie.fr,
	Laboratoire Physico-Chimie, Institut Curie, CNRS UMR 168, 26 Rue d'Ulm, 
	75005, Paris France, Tel.:~+33 (0)1 56 24 67 54. }  \\
	Laboratoire Physico-Chimie,\\
	 Institut Curie, CNRS UMR 168, Université Pierre et Marie Curie-Paris 6\\
	  Paris France}

\date{}

\begin{document}

% generate the title page from the info in the headers above
\maketitle

% 200 words max Abstract
\abstract{Protein mobility is conventionally analyzed in terms of an effective diffusion. Yet, this description often fails to properly distinguish and evaluate the physical parameters (such as the membrane friction) and the biochemical interactions governing the motion. Here, we present a method combining high-density single-molecule imaging and statistical inference to separately map the diffusion and energy landscapes of membrane proteins across the cell surface at  $\sim100$ nm resolution (with acquisition of a few minutes). When applying these analytical tools to glycine neurotransmitter receptors (GlyRs) at inhibitory synapses, we find that gephyrin scaffolds act as shallow energy traps ($\sim$3 $k_BT$) for GlyRs, with a depth modulated by the biochemical properties of the receptor-gephyrin interaction loop. In turn, the inferred maps can be used to simulate the dynamics of proteins in the membrane, from the level of individual receptors to that of the population, and thereby, to model the stochastic fluctuations of physiological parameters (such as the number of receptors at synapses). Overall, our approach provides a powerful and comprehensive framework with which to analyze biochemical interactions in living cells and to decipher the multi-scale dynamics of biomolecules in complex cellular environments.

\emph{Key words:} Single Molecule, Neurons, Statistical Physics}

% New page
\clearpage

\section*{Introduction}

Determining the parameters that regulate the mobility of proteins in cells is key for many cellular functions. The motion of proteins depends on a variety of factors, including the local viscosity, their intermittent binding to other proteins,  the molecular crowding and the dimensionality of the accessible space \cite{Saxton1997}. Since all these factors are difficult or impossible to reconstitute \emph{in vitro} using purified constituents, there is a compelling need for analytical tools that bypass \emph{in vitro} assays and directly access the properties of macromolecular assemblies and the kinetics of their interactions in their native cellular environment. 

Thanks to single-molecule (SM) imaging tools, it is now possible to record trajectories of individual proteins in a variety of cellular systems. An important challenge is to extract relevant biochemical and biophysical information from these trajectories. This is commonly done by computing the mean square displacement (MSD) along the trajectories and estimating the effective diffusion coefficient of the molecule.  By associating the diffusional states to the functional states of the biomolecules, one can identify molecular behaviors \cite{Saxton1997} and evaluate the transition kinetics between them \cite{Persson2013}. Although this approach has often proved useful, it is conceptually inappropriate in many biological situations. Measuring a diffusion coefficient places emphasis on the friction encountered by the protein and assumes that the movement is characterized by a MSD scaling linearly with time. Yet, the primary factor controlling the motion of a protein is often not the friction but, rather, its interactions with molecular or macromolecular partners leading to transient stabilization or transport. In this case, the relevant information is not the diffusion coefficient but the binding energies between the protein of interest and its interacting partners. Furthermore, regulatory processes are often mediated by changes in these binding energies, which should ideally be evaluated with \emph{in situ} measurements. 

Methods that go beyond the computation of the MSD generally aim to identify deviations from Brownian movement within single molecule trajectories, due for instance to trapping or transport \cite{Simson1995,Huet2006,Bouzigues2007}. However these methods essentially remain \textit{ad hoc} tools and do not constitute a comprehensive framework to describe the parameters underlying the motion. Furthermore, biological media are often spatially inhomogeneous and this heterogeneity is poorly conveyed by measuring a few, sparse trajectories. A conceptually different approach using Bayesian inference methods has been recently proposed to analyze the motion of molecules \cite{Masson2009,Turkcan2012}.  It assumes that the  membrane environment is characterized by two spatially-varying quantities: (i) the diffusivity $D\left(\mathbf{r}\right)= k_BT/\gamma\left(\bf{r}\right)$ (where $\gamma\left(\bf{r}\right)$ is the local viscosity), (ii) the potential energy $V\left(\bf{r}\right)$ that reflects the biochemical interactions of the molecule. In this framework, the protein is a random walker with a motion governed by the Langevin equation \cite{Risken1997}:
\begin{equation}
\frac{d\mathbf{r}}{dt}= -\frac{D\left(\mathbf{r}\right) \nabla V\left(\mathbf{r}\right)}{k_BT}+\sqrt{2D\left(\mathbf{r}\right)}\xi\left(t\right)
\label{langevin}
\end{equation}
(\noindent where $\xi\left(t\right)$ is a rapidly varying Gaussian noise with zero mean). From a general standpoint, a knowledge of $D\left(\bf{r}\right)$ and $V\left(\bf{r}\right)$, which are protein-specific, can not only reveal how fast the protein moves in the membrane but also identify areas where it can be stabilized (energy traps) or from which it is excluded (energy barriers). However, in the few cases where $D\left(\bf{r}\right)$ and $V\left(\bf{r}\right)$ have been experimentally determined \cite{Turkcan2013,Hoze2012}, the analysis has been limited to movements confined in local regions ($<$ 1 $\mu$m$^{2}$), falling short of providing a complete description of the heterogeneous diffusivity and energy landscapes in the cell membrane. 

Here, we  introduce a novel and generic approach, combining high density single molecule imaging and computational tools, that enables the mapping of the environment of membrane receptors across the entire cell surface and at $\sim$ 100 nm resolution.  This approach allows the mapping of the membrane over regions of several hundred $\mu m^{2}$ in a few minutes of data acquisition. Furthermore, the inferred maps are used to numerically generate massive number of trajectories. These simulated trajectories, whose characteristics match those of the experimental ones, enable a complete analysis of the dynamics in the complex membrane environment by means of various statistical estimators. To illustrate the relevance and benefits of our approach, we applied it to the neuronal membrane, a cellular system in which the spatial organization is critical for the detection and processing of external information. In past years, tracking experiments have underlined the role of membrane dynamics in ensuring rapid exchange of receptors (e.g. glutamate, glycine or GABA receptors) between extrasynaptic and synaptic localizations \cite{Triller2008}. Therefore, the number of receptors at synapses depends on the motion of receptors at the cell surface and their stabilization at synaptic loci, the latter being regulated by the number of scaffolding molecules and the affinity of the receptor-scaffold interactions \cite{Renner2009}. A quantitative analysis of the protein mobilities and of their regulatory mechanisms is thus paramount for characterizing and modeling the variability of the synaptic response and the plasticity of the nervous system (involved in higher brain functions such as learning and memory or during pathological processes). 

\section*{Results and discussion}

\begin{figure}
   \begin{center}
      \includegraphics*[width=3.25in]{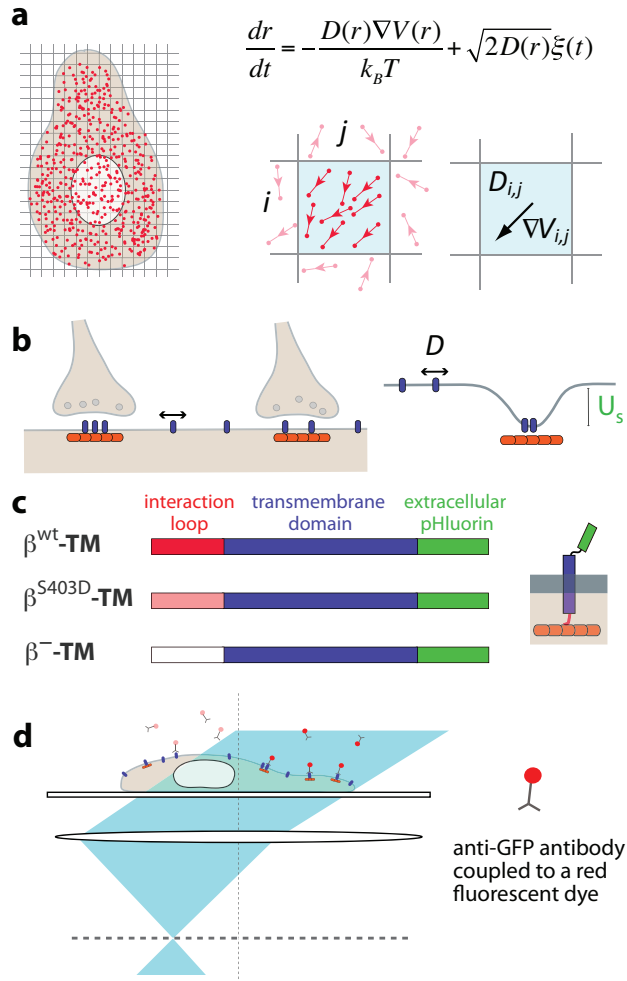}
      \caption{ General scheme of the assay. a) Principle of the Bayesian inference method. Left: High-density single-molecule data (red dots) are recorded at the cell surface. Right: In a mesh domain, multiple translocations (top) are used to infer the local diffusivity and force (gradient of the potential) that underlie the motion (bottom). b)  GlyRs (blue) diffuse in the membrane and are in dynamic equilibrium between synaptic and extrasynaptic domains in the neuronal membrane. At synapses, GlyRs are stabilized by their interactions with gephyrin clusters (orange), which can be modeled as trapping potential (with depth $U_S$). c) Expression constructs of transmembrane (TM) proteins with an extracellular pHluorin tag and an intracellular interaction loop derived from the GlyR $\beta$-subunit. d) Principle of high-density single-molecule uPaint imaging \cite{Giannone2010}.   }
      \label{fig:result_fig}
   \end{center}
\end{figure}

\noindent \textbf{Mapping the diffusion and energy landscapes with Bayesian inference}. Our approach for the large-scale mapping of $D\left(\bf{r}\right)$ and $V\left(\bf{r}\right)$ builds on Bayesian statistical tools recently developed to analyze the motion of individual particles \cite{Masson2009,Turkcan2012}. The principle of the method is as follows (see details in the Supplementary materials). We first acquire high-density single-molecule data \cite{Manley2008,Giannone2010}, with a number of individual translocations on the order of $1000-10 000\, / \mu$m$^2$. Next, the surface of the cell is meshed with sub-domains $S_{i,j}$ (labeled with the index $(i,j)$ along the $x$ and $y$ axis)  with a size proportional by a factor $\delta\sim$ 2-3 to the average step size of a translocation, such that consecutive positions of the molecules are either in the same or in adjacent domains (Fig. 1a). From the information contained in the massive number of individual translocations, we determine $D_{i,j}$ and $\nabla V_{i,j}$ in each sub-domain $(i,j)$ using Bayesian inference techniques adapted from \cite{Masson2009}. In brief, we compute the global posterior distribution $P$ of the parameters $\lbrace{D_{i,j}\rbrace}_{\left( i,j\right)}$ and $\lbrace{\nabla V_{i,j}\rbrace}_{\left( i,j\right)}$ given the observed trajectories $\lbrace{T_{k}\rbrace}_{\left(k\right)}$. Since all the sub-domains are independent, $P$ is the product of the posterior distributions inside each of them:
%\begin{equation}
\begin{eqnarray}
 %P\left(\lbrace{  \nabla V_{i,j}  \rbrace}_{\left( i,j\right)},\lbrace{D_{i,j}\rbrace}_{\left( i,j\right)}|\lbrace{T_{k}\rbrace}_{\left(k\right)} \right) &=&  \left( \prod_{\left(i,j\right)}P\left( \nabla V_{i,j} , D_{i,j}|\lbrace{T_{k}\rbrace}_{\left(k\right)} \right) \right) \textcolor{blue}{ \times P\left(\nabla V, D \right) }\\
 P\left(\lbrace{  \nabla V_{i,j}  \rbrace}_{\left( i,j\right)},\lbrace{D_{i,j}\rbrace}_{\left( i,j\right)}|\lbrace{T_{k}\rbrace}_{\left(k\right)} \right) =  \left( \prod_{\left(i,j\right)}P\left( \nabla V_{i,j} , D_{i,j}|\lbrace{T_{k}\rbrace}_{\left(k\right)} \right) \right)  \times P\left(\nabla V, D \right) \\
 %&\propto& \prod_{\left(i,j\right)}\left( \prod_{k} \prod_{\mu:\bf{r}^{k}_{\mu}\in S_{i,j}}  \frac{  \exp\left(-\frac{  \left( \bf{r}^{k}_{\mu + 1} - \bf{r}^{k}_{\mu} - D_{i,j}\nabla V_{i,j}\Delta t/k_{B}T \right)^{2}    }{ 4\left(D_{i,j}+\frac{\sigma^{2}}{\Delta t}\right)\Delta t     }      \right)               }{4\pi\left(D_{i,j}+\frac{\sigma^{2}}{\Delta t}\right)\Delta t }   \textcolor{blue}{ \times P\left(\nabla V, D \right) }  \right) 
\propto \prod_{\left(i,j\right)}\left( \prod_{k} \prod_{\mu:\bf{r}^{k}_{\mu}\in S_{i,j}}  \frac{  \exp\left(-\frac{  \left( \bf{r}^{k}_{\mu + 1} - \bf{r}^{k}_{\mu} - D_{i,j}\nabla V_{i,j}\Delta t/k_{B}T \right)^{2}    }{ 4\left(D_{i,j}+\frac{\sigma^{2}}{\Delta t}\right)\Delta t     }      \right)               }{4\pi\left(D_{i,j}+\frac{\sigma^{2}}{\Delta t}\right)\Delta t } \times  \frac{D^{2}_{i,j}}{\left( D_{i,j}\Delta t + \sigma^2 \right)^{2}}   \right) 
\label{posterior}
\end{eqnarray}
%\end{equation}
\noindent  where $\mu$ designates the index for which the points $\bf{r}^{k}_{\mu}$ of the $k^\mathrm{th}$ trajectory are in $S_{i,j}$, $\sigma$ is the experimental localization accuracy ($\sim$30 nm), $\Delta t$ the acquisition time and $P\left(\nabla V, D \right)$ the prior information on the potential and the diffusivities.  In the second line of equation (2) we display the prior we commonly used, Jefferey's prior, that is discussed in the supplementary Materials. The estimators  ($D_{i,j}^\mathrm{MAP}$, $\nabla V_{i,j}^\mathrm{MAP}$) of the local diffusivity and force are the Maximum a Posteriori (MAP) of the posterior distribution $P$ \cite{MacKay2003,vonToussaint2011}. Finally, we solve the inverse problem to determine in each sub-domain the potential field $V_{i,j}$ associated to the force. The estimation of $V_{i,j}$ is performed by minimizing $\xi\left( \lbrace{V_{i,j}\rbrace} \right)$, defined as:
\begin{equation}
\xi\left(\lbrace{V_{i,j}\rbrace}|\left(i,j\right)\in\lbrace{N\left(i,j\right)\rbrace}\neq0\right)= \\
\sum_{\left(i,j\right)}\left(\nabla V_{i,j}-\nabla V^{MAP}_{i,j}\right)^{2}+\beta\left(\delta\right)\sum_{\left(i,j\right)}\left(\nabla V_{i,j}\right)^{2}
\label{minimize}
\end{equation}
with $N\left(i,j\right)$ the number of neighboring occupied mesh domains, $\beta\left(\delta\right)$ a constant (optimized on numerically generated trajectories) depending on $\delta$ (see Supplementary Information). Eventually, the set of quantities $\lbrace{D_{i,j}^\mathrm{MAP}, \nabla V_{i,j}\rbrace}_{\left( i,j\right)}$ constitute the diffusivity and potential energy maps.

\noindent \textbf{Glycine receptors and their interactions with scaffolding proteins}. We applied our inference-based mapping method to investigate the dynamics of GlyRs in the neuronal membrane as well as their stabilization at inhibitory synapses \cite{Dahan2003}. This stabilization is achieved through the binding of the receptors to the scaffold protein gephyrin (Fig. 1b) via an intracellular loop (the $\beta$-loop) present in the two $\beta$-subunits of the pentameric GlyR complex. The high affinity component of the $\beta$-loop-gephyrin interaction is in the nanomolar range ($K_{D} \sim$ 20 nM), as determined by isothermal titration calorimetry \cite{Specht2011}. To characterize the GlyR-gephyrin interaction in living neurons, we used recombinant membrane proteins consisting of a transmembrane domain (TM), a C-terminal pHluorin tag (a pH-sensitive GFP mutant that is quenched in intracellular acidic vesicular compartments), that were fused N-terminally to the intracellular GlyR $\beta$-loop (Fig. 1c). This $\beta^{\mathrm{WT}}$-TM-pHluorin construct recapitulates the interactions of the endogenous GlyR complexes with the gephyrin scaffold proteins, with the important benefit that individual elements of the receptor-scaffold interaction can be manipulated independently \cite{Specht2011}. It also overcomes the difficulty of defining the sub-unit composition of oligomeric receptors where transfected sub-units compete with endogenous ones. As a control, we used $\beta^{-}$-TM-pHluorin, a construct with a mutated $\beta$-loop that does not interact with gephyrin. \\

\noindent \textbf{High-density single-molecule imaging of TM proteins}. We acquired a high-density of individual trajectories using uPaint, a single-molecule technique in which cells are imaged at an oblique illumination in a buffer containing dye-labeled primary antibodies  \cite{Giannone2010}. As antibodies (in our case, anti-GFP antibodies coupled to Atto647N dyes) continuously bind to their membrane targets, they can be tracked until they either dissociate or photo-bleach (Fig. 1d and Supplementary Videos 1 and 2). Hence, the entire field of view is constantly replenished with new fluorescent labels and a large number of individual trajectories covering a field of view of $\sim$ 500-1000 $\mu m^{2}$ can be recorded. Experiments were performed on cultured rat hippocampal neurons co-transfected with mRFP-tagged gephyrin and with the pHluorin-tagged transmembrane constructs (Fig. 1c). In typical measurements, movies were recorded for $\sim$5-15 minutes with an acquisition time $\Delta t$  = 50 ms (Supplementary Videos 1 and 2), yielding up to hundreds of thousands of individual translocations per field of view, with an average of 30 points per mesh domain (size $\sim$100x100 nm$^2$). On this time scale, the cells and synaptic sites remained relatively stable, meaning that the diffusivity and energy landscapes could be considered constant.  \\

\noindent \textbf{Diffusion and energy maps of TM proteins}. Figures 2a-f show examples of the diffusivity and energy maps for the two constructs $\beta^{\mathrm{WT}}$-TM-pHluorin and $\beta^{-}$-TM-pHluorin. In both cases, the diffusion map exhibits fluctuations at short scale ($\sim$1 $\mu$m or less), with local peaks and valleys and a characteristic diffusivity in the range of 0.05-0.2 $\mu$m$^{2}$.s$^{-1}$ (Fig. 2b and 2e). More striking differences were observed between the energy landscapes. For $\beta^{\mathrm{WT}}$-TM, the landscape is characterized by the existence of small regions ($<$ 0.5 $\mu$m$^2$) corresponding to local energy minima (Fig. 2c). Importantly, gephyrin clusters coincide with energy minima, consistent with the stabilization of the transmembrane proteins at synaptic sites. Yet, we also observed that some other minima did not colocalize with gephyrin clusters, suggesting that $\beta^{\mathrm{WT}}$-TM-pHluorin might interact with other partners outside of synapses (such as the cytoskeleton or lipid domains). It is possible that these extrasynaptic interactions are still mediated by gephyrin (present in number too small to be detected), since gephyrin is known to associate with GlyRs both inside and outside of synapses \cite{Ehrensperger2007}. In contrast, the energy map for  $\beta^{-}$-TM (Fig. 2f) shows variations at a longer length-scale, without correlation to gephyrin clusters.  \\

\begin{figure}
   \begin{center}
      \includegraphics*[width=3.25in]{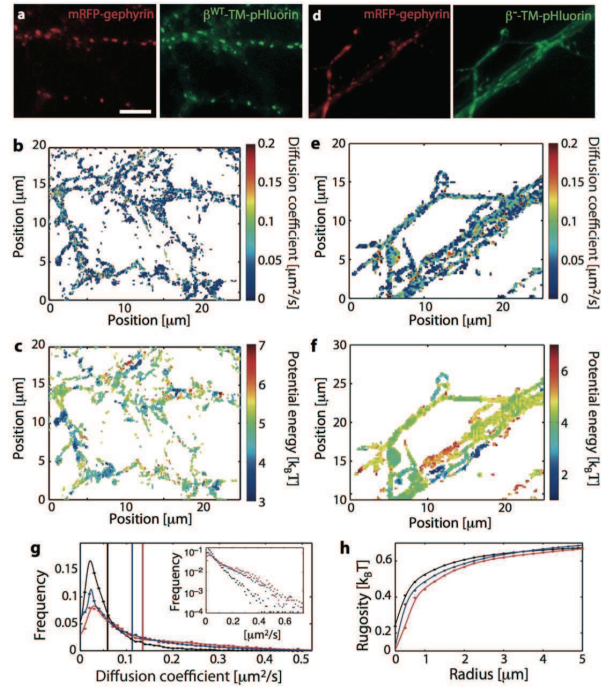}
      \caption{Diffusion and energy maps in live neurons. a) Fluorescence images of cultured neurons expressing mRFP-gephyrin and $\beta^{\mathrm{WT}}$-TM-pHluorin. Scale bar: 10 $\mu$m. b-c) Diffusion and energy maps. d-f) Equivalent set of images and maps for $\beta^{-}$-TM-pHluorin. g) Distribution of diffusion coefficients for the membrane constructs $\beta^{\mathrm{WT}}$-TM (black), $\beta^{S403D}$-TM (blue) and $\beta^{-}$-TM (red). The vertical bars on the x-axis indicate the mean values of the respective distributions. Insert: distribution in a lin-log scale. h) Rugosity of the membrane potential as a function of the region radius.}
      \label{fig:result_fig}
   \end{center}
\end{figure}

To more quantitatively compare the heterogeneous properties of the neuronal membrane for $\beta^{\mathrm{WT}}$-TM and $\beta^{-}$-TM, we computed two quantities (averaged over 7 cells in each case): (i) the distribution of diffusion coefficients in the maps (Fig. 2g), and (ii) the rugosity of the energy landscape, defined as the standard deviation of the potential in circular region of defined radius averaged over the complete surface of the cell over circular region of the  fluctuations of the potential averaged over circular regions of increasing radius (Fig. 2h and Supplementary Information). These parameters revealed that the interacting $\beta$-loop led to a lower average diffusivity (0.06 $\mu$m$^{2}$.s$^{-1}$ and 0.13 $\mu $m$^{2}$.s$^{-1}$ for $\beta^{\mathrm{WT}}$-TM and $\beta^{-}$-TM, respectively) and a larger rugosity of the potential. This is consistent with the notion that moving TM proteins, when bound to intracellular scaffolding proteins, encounter more obstacles that increase the viscosity of their environment. Also they are more likely to interact with membrane or sub-membrane structures that contribute to the energy landscapes.\\

\noindent \textbf{Synaptic scaffolds as crowded energy traps}. Given the pronounced differences between the energy landscapes of the $\beta^{\mathrm{WT}}$-TM and $\beta^{-}$-TM constructs, we examined the behavior of $\beta^{\mathrm{WT}}$-TM at gephyrin clusters in closer details. An example of the energy profile of $\beta^{\mathrm{WT}}$-TM proteins at a synaptic cluster (identified by the presence of mRFP-gephyrin fluorescence) is shown in Fig 3a. The profile reinforces the view that clusters of scaffolding proteins act as energy traps for membrane receptors \cite{Hoze2012,Triller2008,Dahan2003}. The average trap depth was 3.6 $\pm$ 0.4 $k_BT$ ($\pm$ S.E.M., $n$ = 69 clusters), a relatively shallow potential from which receptors can escape rapidly. Yet, about 15\% of clusters had stabilization energies greater than 6 $k_BT$, corresponding to a much more stable anchoring of receptors (Fig. 3b). This reflects the heterogeneity of the synaptic domains in the neuronal membrane and underlines the need for measurements at the single synapse level.
 Of note, the binding energies between $\beta^{\mathrm{WT}}$-TM and gephyrin seem to be significantly lower than the stabilization energy of AMPA receptors at synaptic sites, for which 25 $\%$ of the wells had a depth larger than 8 $K_BT$ \cite{Hoze2012}. The method used in \cite{Hoze2012}, also based on a combination of high-density single molecule imaging and statistical inference, evaluates the diffusion and drift by computing the maximal likelihood estimation in a mesh square as described in \cite{Masson2009}. The confining potentials were subsequently evaluated by L2 minimization of a parabolic shaped potential from the force (drift) fields. In \cite{Hoze2012} the authors do not discuss the role of known biases with confining potentials (see \cite{Turkcan2012,Turkcan2013,Voisinne2010}) or the effect of the positioning noise, and do not provide information on the posterior distribution of the parameters. It is thus delicate to precisely compare their experimental results with ours. Yet, given that the diffusivity of AMPARs at excitatory synapses appears to be higher than the diffusivity of GlyRs at inhibitory synapses (gephyrin clusters), higher confining potentials may be necessary to stabilize the AMPARs. In addition, we noticed that the average diffusivity of $\beta^{\mathrm{WT}}$-TM ($\sim$0.01 $\mu m^{2}s^{-1}$) inside gephyrin clusters was reduced by a factor $\sim$6 compared to extrasynaptic regions (Fig. 3c), probably due to the combined effect of membrane crowding within synaptic sites and the binding to scaffolding elements. In comparison, the diffusivity of $\beta^{-}$-TM proteins inside gephyrin clusters, which we expect to be predominantly influenced by molecular crowding \cite{Renner2009b}, was 0.07 $\mu$m$^{2}$.s$^{-1}$  (Fig. 3c), only a factor $\sim$2 lower than in extrasynaptic domains. In other words, the synaptic scaffold stabilizes the receptor by simultaneously diminishing the diffusivity of the receptor and by acting as a trapping potential. \\

\begin{figure}
   \begin{center}
      \includegraphics*[width=3.25in]{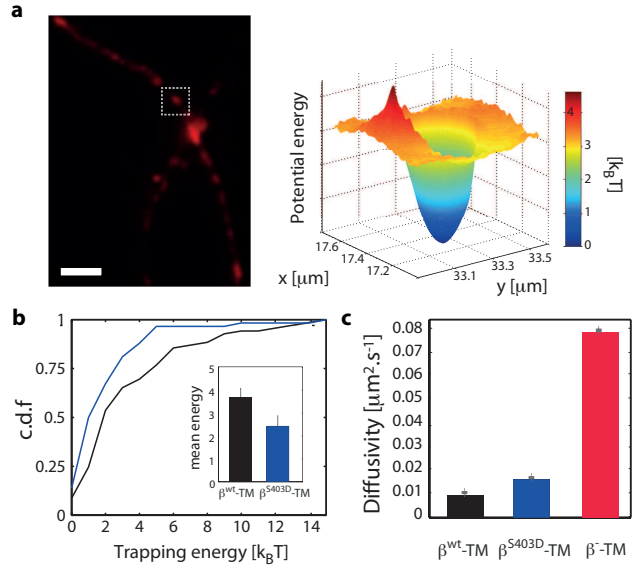}
      \caption{Analysis of the synaptic gephyrin scaffold. a) Example of a gephyrin cluster (indicated by a box) acting as a local trap in the energy landscape. Scale bar: 5 $\mu$m b) Cumulative distribution function (c.d.f) of trapping energy for the constructs  $\beta^{\mathrm{WT}}$-TM (black) and $\beta^{S403D}$-TM (blue). Insert: mean values of the distribution. Error bars indicate the S.E.M. c) Mean diffusivity for $\beta^{\mathrm{WT}}$-TM (black), $\beta^{S403D}$-TM (blue) and $\beta^{-}$-TM (red). Error bars indicate the S.E.M.}
      \label{fig:result_fig}
   \end{center}
\end{figure}

\noindent \textbf{Modulation of the $\beta$-loop gephyrin binding affinity}. Since the computation of the energy landscape allows the unambiguous distinction between interacting membrane constructs and those lacking interaction domains, we tested the sensitivity of our approach with the phosphomimetic construct $\beta^{S403D}$-TM, a mutated $\beta$-loop  known to have a lower gephyrin binding affinity \emph{in vitro} ($K_{D}$ $\sim$ 0.9 $\mu$M \cite{Specht2011})  (Fig. 1c). As a result, $\beta^{S403D}$-TM displayed increased membrane diffusion and reduced synaptic accumulation compared to $\beta^{\mathrm{WT}}$-TM. The phosphorylation of amino acid residue S403 of the GlyR$\beta$ subunit by protein kinase C thus contributes to the regulation of GlyR levels at inhibitory synapses \cite{Specht2011}. The diffusion and energy landscapes of $\beta^{S403D}$-TM (computed over 6 different cells) yielded a diffusivity (average value 0.11 $\mu$m$^{2}$.s$^{-1}$) and an energetic rugosity precisely intermediate between those of the wild-type and of the binding-deficient constructs (Fig. 2g-h). Compared to $\beta^{\mathrm{WT}}$-TM, the average trap depth of $\beta^{S403D}$-TM at synaptic sites was reduced to 2.4 $\pm$ 0.4 $k_BT$ ($n$ = 58 clusters), with less than 5\% of the traps above 6 $k_BT$ (Fig. 3b). Inside clusters, the average diffusivity (0.015 $\mu$m$^{2}$.s$^{-1}$) was slightly higher than for the wild-type (Fig. 3c).

Importantly, the binding energy reported here corresponds to TM proteins moving in a two-dimensional membrane and interacting with macromolecular gephyrin scaffolds that are believed to be two-dimensional as well \cite{Fritschy2008,Specht2013}. This is in contrast with measurement of the equilibrium constant $K_{D}$ by isothermal calorimetry, which reports on the individual interaction between the $\beta$-loop and the scaffolding protein in an isotropic, three-dimensional measurement of the $\beta$-loop-scaffold interaction. Obtaining the stabilization energy thus constitutes a first and important step to bridge the gap between \emph{in vitro} and \emph{in situ} biochemical measurements. When further complemented with data on the ultra-structure and stoichiometry of synaptic scaffolds (that are now accessible with single molecule imaging techniques \cite{Specht2013,Lord2010}), we expect our approach to enable a true determination of the two-dimensional affinity of the membrane proteins for the synaptic scaffolds \cite{Wu2011}. \\

\noindent \textbf{Connecting the landscapes and the global mobility of proteins}. An important question for the dynamics of proteins is how the variability of their diffusion and energy landscapes at short scale ($\sim$ 100 nm) affects their long-distance mobility and, thereby, the kinetics of many intermolecular reactions. Reaching a multi-scale description of the motion in the membrane has long been a challenge in single-molecule experiments. High-density sampling is usually achieved with poorly stable probes, yielding numerous but short trajectories \cite{Giannone2010,Manley2008}. In contrast, long trajectories obtained with more stable markers such as quantum dots \cite{Pinaud2010} only provide a sparse sampling of the cell surface. Furthermore, the nature of the motion, such as sub-diffusion, may prevent efficient space sampling with single long trajectories. Here, we adopted a different strategy and used the inferred maps as phenomenological templates to simulate the motion of proteins. Practically, we used the Gillespie scheme \cite{Gillespie1977} to generate individual trajectories lasting up to 500s (see Methods and Supplementary Information).

\begin{figure}
   \begin{center}
      \includegraphics*[width=3.5in]{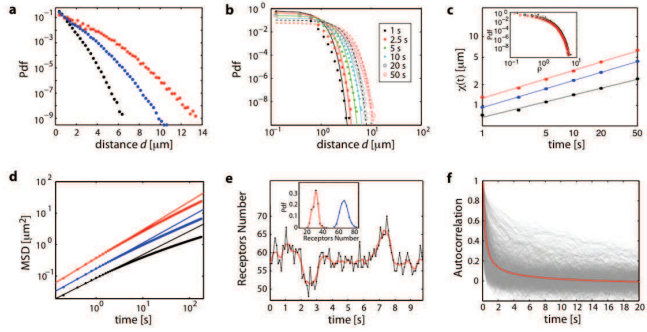}
      \caption{Analysis of simulated trajectories in the inferred maps. Unless otherwise mentioned, the results correspond to the constructs $\beta^{\mathrm{WT}}$-TM (black), $\beta^{S403D}$-TM (blue) and $\beta^{-}$-TM (red).  a) Ensemble-averaged propagator $\Pi(d,t)$, defined as the probability density function (pdf) to move by a given distance in $t$=10 s. b) Propagator $\Pi(d,t)$ for the construct $\beta^{\mathrm{WT}}$-TM, computed at different times $t$. The plain lines are adjustments with the gaussian curves $\exp(-d^2/2\chi^2(t))/2\pi\chi^2(t)$. c) Curves $\chi(t)$. Inset: Propagators for the construct $\beta^{\mathrm{WT}}$-TM as a function of the rescaled variable $\rho = d/\chi(t)$. d) Mean-squared displacement (MSD) as a function of time. The straight lines indicate the sub-diffusive behavior at short time-scales. e) Time-course of the number of receptors at a single synapse. Inset: distribution of the minimum (in red) and maximum (in blue) number of receptors computed over traces of 300 s for all the gephyrin clusters. f) Autocorrelation functions (in grey) for the time traces of number of receptors at gephyrin clusters (computed over 300 s). The red line indicates the average autocorrelation function.}
      \label{fig:result_fig}
   \end{center}
\end{figure}

From a large number of simulated trajectories, we could compute ensemble-averaged quantities. We first evaluated the propagator $\Pi(d,t)$, namely the probability density function of moving a distance $d$ in a time $t$, which is the fundamental estimator characterizing the random motion in a complex environment \cite{Metzler2004}. Although the difference in the average trapping energy at gephyrin clusters was only $\sim$1 $k_BT$  between $\beta^{\mathrm{WT}}$-TM and $\beta^{S403D}$-TM, it led to significant changes in the mobility, reducing the probability of moving over long distances with increasing strength of the $\beta$-loop-gephyrin interaction (Fig. 4a). To more carefully examine the nature of the movement of $\beta^{\mathrm{WT}}$-TM, we plotted $\Pi(d,t)$ at different times $t$. The curves could be approximated by gaussian curves $\exp(-d^2/2\chi^2(t))$ with $\chi(t) \propto t^{\alpha}$ and $\alpha = 0.33$, less than 0.5 the value expected for a standard Brownian motion (Fig. 4b-c). In fact, these results are consistent with a subdiffusive movement resulting from a fractional brownian motion due to heteregoneities in the diffusion and energy landscapes \cite{Metzler2004}. Similar results were obtained for $\beta^{S403D}$-TM and $\beta^{-}$-TM, with $\alpha$ increasing to 0.39 and 0.41, respectively (Fig. 4c). The subdiffusive nature of the motion could be further illustrated by computing the ensemble-averaged MSD for the three transmembrane constructs (Fig. 4d). On the time scales 0.05-5 s, all the MSDs increased sublinearly, with an anomalous exponent $\alpha$ equal to 0.75, 0.82 and 0.89 for $\beta^{\mathrm{WT}}$-TM, $\beta^{S403D} $-TM and $\beta^{-}$-TM, respectively. The MSD anomalous exponents are slightly larger than 2$\alpha$, likely due to boundary effects associated to the size and geometry of the neurons.

Finally, we examined the implications of the local properties of the mobility of individual GlyRs on their global distribution in the membrane and on the receptor occupancy at synapses.
To do so, we simulated the membrane dynamics of a population of receptors, using surface densities derived from prior experimental reports (Supplementary information). We computed in particular the time course of the number of receptors at individual synaptic clusters, which we expect to fluctuate due to the exit and entry fluxes of receptors (Fig. 4e and Supplementary Information). The exit kinetics at a given synapse is determined by the shape and amplitude of the trapping potential combined with the reduced diffusivity in the cluster. In contrast, the entry kinetics depends on the motion of all the receptors over the entire cell surface and need to be computed using the diffusion and energy maps. The number of receptors varied significantly over times, as illustrated by the distribution of their minimal and maximal numbers at individual synapses (Fig. 4e). Furthermore, the time scale of these fluctuations, analyzed by computing the autocorrelation function, is comprised between $\sim$ 1 s and a few tens of seconds, showing a large heterogeneity among gephyrin clusters (Fig. 4f). These observations may account for the dynamic range of receptor numbers at synapses and for the variability of synaptic transmission \cite{Ribrault2011}. The receptor fluctuations, that are equivalent to a noise, may also favor the transition from one steady state to another during synaptic plasticity \cite{Sekimoto2009}.

\section*{Conclusion}

The motion of proteins in the plasma membrane is influenced by both a viscous landscape, $\gamma\left(\bf{r}\right)$, and an interaction potential, $V\left(\bf{r}\right)$.  We have introduced a method to map the interaction energy and diffusion landscapes in the cellular membrane with $\sim$ 100 nm resolution over surfaces of several hundred $\mu m^{2}$. The possibility of simulating trajectories in the inferred maps offers many possibilities to address the multiscale dynamics of membrane proteins. In particular, it bridges the gap between the information obtained from numerous, dense – but short - trajectories acquired using uPaint or sptPALM techniques, and that from the much longer, but usually sparse, trajectories extracted through the tracking of proteins labeled with photostable fluorophores (Qdots, nanoparticles).
These trajectories can be used to accurately evaluate various statistical estimators, thus enabling the analysis of the dynamics of biomolecules in complex media. We anticipate that our method will be instrumental to identify the factors  governing the mobility of specific molecules (such as friction, molecular interactions and geometry of the cell) and thereby to model and analyze reaction-diffusion processes in biological media. As illustrated in the case of GlyR-gephyrin binding, it also paves the way to \emph{in situ} biochemical measurements, which is key for a quantitative analysis of the regulation of molecular interactions in a cellular environment. Our approach should also be helpful to describe the molecular noise that results from variability of protein concentrations across the cell surface and may play an important role in information processing at the single cell level \cite{Ribrault2011}. Beyond the case of receptor-scaffold interactions, our analytical tools can be applied to other biological questions, such as the stability of macromolecular assemblies in the cytoplasm or the nucleus, or to the sequence-dependent movement of proteins along DNA \cite{Leith2012}. 

 \section*{Materials and methods}
 
\noindent \textbf{Antibody coupling.}
Rat anti-GFP monoclonal antibody (Roche) was labeled with Atto-647 dye using standard conjugation methods. In brief, 40 $\mu$L of antibodies at 0.4 mg/ml in PBS were mixed with 4 $\mu$L of 1 M sodium bicarbonate buffer at pH 8.5. This solution was incubated with 10-fold molar excess of Atto-647-NHS-ester (Sigma) diluted at 1 mg/mL in anhydrous DMSO. After 1 h of incubation at room temperature, the solution was filtered with a Microspin G50 column (GE Healthcare) to remove unconjugated dye. The overall coupling efficiency of the dye, estimated by UV-Vis absorption, was about 12\%. The labeled antibodies were washed with PBS and concentrated using three rounds of centrifugation with a vivaspin500 10 kDa cut-off PES membrane filter (GE Healthcare). The concentrated antibody solution was stored at 4 C and used for up to one week. \\

\noindent \textbf{Cell culture and plasmid transfection.} 
Hippocampal neurons from Sprague Dawley rats at embryonic day 18 were cultured at a density of 6 x $10^{4}$ cells/cm$^2$ on 18 mm coverslips precoated with 80 mg/ml poly-D,L-ornithine (Sigma) and 5\% fetal calf serum (Invitrogen) as described previously \cite{Specht2011}. Cultures were maintained in serum-free Neurobasal medium supplemented with 1x B27 and 2 mM glutamine (Invitrogen). Cells were transfected after 6 to 8 days in vitro using Lipofectamine 2000 (Invitrogen), and imaged 1 to 2 days after transfection. All coverslips were co-transfected with mRFP-tagged gephyrin and pHluorin-tagged TM constructs, using 0.4 $\mu$g of each plasmid per coverslip. The expression constructs $\beta^{\mathrm{WT}}$-TM-pHluorin, $\beta^{S403D}$-TM-pHluorin and $\beta^{-}$-TM-pHluorin are all described in \cite{Specht2011}. In brief, $\beta^{S403D}$ corresponds to the mutation of serine S403 of the GlyR$\beta$ subunit that mimics the phosphorylation of the residue by protein kinase C. $\beta^{-}$-TM corresponds to the double mutation F398A and I400A of the wild-type GlyR $\beta$-loop that abolishes binding to gephyrin. \\

\noindent \textbf{Cell labeling.}
Prior to imaging, we prepared a stock solution of diluted antibodies using casein (Vector laboratories) as a blocking reagent. We added 2 $\mu$l of Atto-647 conjugated anti-GFP antibodies and 10 $\mu$l of 10 mg/ml casein to 40 $\mu$l of PBS, resulting in an antibody solution of 0.1-0.2 $\mu$M. We also prepared a stock of Tetraspeck fluorescent microbeads (Invitrogen) by mixing 1 $\mu$l of 0.1 $\mu$M microbeads with 400 $\mu$l of imaging solution. These multi-color fluorescent beads were used as a reference to align the different imaging channels and to correct for $x/y$ drifts of the stage and the coverslip. The coverslip was mounted in an imaging chamber and incubated with 20 $\mu$l of warmed microbead solution for 10 seconds. After rinsing, the chamber was filled with 600 $\mu$l of warmed imaging solution (MEMair: phenol red-free MEM, glucose 33 mM, HEPES 20 mM, glutamine 2 mM, Na-pyruvate 1 mM, and B27 1x) and placed on the microscope. To avoid saturating the cell membrane with fluorescent antibodies, we first selected a transfected neuron and added the fluorescent antibodies at a final concentration of 0.3-0.6 nM directly before the start of the acquisition. \\

\noindent \textbf{Imaging.}
Measurements were performed on an inverted epi-fluorescence microscope (Olympus IX70) equipped with a 100x 1.45NA oil objective and a back-illuminated electron-multiplying CCD camera (Quantem, Roper). We imaged the neurons at 37 C in MEMair recording medium using a heated stage. For each neuron, we first recorded images of the pHluorin signal of the TM constructs and of mRFP-gephyrin fluorescence, using a UV lamp (Uvico, Rapp Optoelectronic) and standard sets of filters for GFP (excitation 475AF40, dichroic 515DRLP and emission 535AF45) and RFP (excitation 580DF30, dichroic 600DRLP and emission 620DF30). Next we acquired a uPaint movie of the transmembrane proteins labeled with Atto-647-coupled anti-GFP antibodies (20 000 images at 20 frames/s). Atto-647 dyes were excited with a 640 nm laser and their fluorescence was collected through using a 650DRLP dichroic and a 690DF40 emission filter. The laser was tightly focused on the back focal plane of the objective. The angle of incidence of the beam on the coverslip, controlled by laterally moving the focused spot, was just under the limit of total internal reflection, such that the laser beam in the sample was almost parallel to the glass surface. This angle was slightly adjusted in each experiment to maximize the signal/noise ratio of the single fluorescent spots diffusing in the membrane. \\

\noindent \textbf{Data analysis.}
Tracking analysis of the movies was carried out using an adapted version of the multiple target tracking algorithm \cite{Serge2008}. In brief, fluorescence spots corresponding to the point-spread function of single emitting fluorophores were fitted with a two-dimensional Gaussian. The centre of the fit yielded the position of single molecules with localization accuracy $\sim$ 30 nm. Trajectories were then computed from individual detections with a nearest-neighbor algorithm. \\

\noindent \textbf{Simulations in the landscapes.} 
The maps of the diffusion and energy landscapes, $D\left(\bf{r}\right)$ and $V\left(\bf{r}\right)$, can be used to simulate the behavior of the molecules at different time and space scales. In each mesh sub-domain $(i,j)$ a diffusivity $D_{i,j}$  is associated with a potential energy value $V_{i,j}$. The dynamics of the molecules are described by the Fokker-Planck equation:
\begin{equation}
\frac{\partial P\left(\bf{r},t|\bf{r}_{0},t\right)}{\partial t} = 
 -\nabla . \left(   -\frac{\nabla V\left(\bf{r}\right)P\left(\bf{r},t|\bf{r}_{0},t\right)}{\gamma\left(\bf{r}\right)}  - \nabla\left(D\left(\bf{r}\right)P\left(\bf{r},t|\bf{r}_{0},t\right)\right)     \right)
\end{equation}
where $P\left(\bf{r},t|\bf{r}_{0},t\right)$ is the conditional transition probability from $(\bf{r}_{0},t_{0})$ to $(\bf{r},t)$. Fokker-Planck equations can always be approximated by Master equations:
\begin{equation}
\frac{dP_{\left(i,j\right)}\left(t\right)}{dt}= \sum_{\left(i',j'\right)\in N\left(i,j\right)}W_{\left(i,j\right),\left(i',j'\right)}P_{\left(i',j'\right)} -
 \sum_{\left(i',j'\right)\in N\left(i,j\right)}W_{\left(i',j'\right),\left(i,j\right)}P_{\left(i,j\right)}
 \label{maitresse}
\end{equation}

with in our case
\begin{equation}
W_{\left(i,j\right),\left(i',j'\right)}= \frac{D\left(i',j'\right)}{\Delta x^{2}} \exp \left(-\frac{\Delta x F^{x}_{\left(i,j\right),\left(i',j'\right)}}{2\gamma\left(i',j'\right)D_{\left(i',j'\right)}}\right)
\end{equation}
if the transition is in the x direction and a similar formula in the y direction and with $W_{\left(i,j\right),\left(i',j'\right)}$ the transition rate from the $\left(i',j'\right)$ site to the $\left(i,j\right)$, $\Delta x$ ($\left(\Delta y\right)$) the mesh size in the x (y) direction, and $F^{x}_{\left(i,j\right),\left(i',j'\right)}$  the potential gradient acting on the random walker in the x direction when moving from $\left(i',j'\right)$ to $\left(i,j\right)$. The motion of the molecule following equation \ref{maitresse} was simulated using the Gillespie scheme \cite{Gillespie1977}. When the molecule was at the site $\left(i,j\right)$, the transitions rates, rewritten $a_{\nu}$ to match Gillespie formalism,$\nu$ taking values from 1 to 4, were evaluated on all neighboring sites. We define $a_{0}=\sum_{\nu}a_{\nu}$. The time, $\tau$, to move from the site $\left(i,j\right)$ to a neighboring site is extracted from an exponential probability density function of rate $a_{0}$, so that $\tau = \frac{1}{a_{0}}\log\left(\frac{1}{r_{1}}\right)$ with $r_{1}$ a random number in $\left[0,1\right]$. The destination site, k, is chosen to satisfy : $\sum^{k-1}_{\nu=0} a_{\nu}\leq r_{2}a_{0}\leq \sum^{k}_{\nu=0} a_{\nu}$ with $r_{2}$ a random number in $\left[0,1\right]$. Limits of the neuronal cells and unvisited sites are defined as inaccessible sites. Note that the trajectory generation process leads to trajectories with non-constant time steps. In order to evaluate the different estimators, trajectories were regularized to obtain the molecule position at regular time lags by imposing that as long as each $\tau$ was not reached the molecule did not move.

\noindent \section*{Acknowledgments}
We are grateful to Paul de Koninck for his support and discussion. We also thank Diego Krapf for his critical reading of the paper and multiple suggestions. This work was funded by CNRS, Inserm, C'Nano Ile de France, the program "Prise de Risque" from CNRS, Agence Nationale pour la Recherche PiriBio, the grant Synaptune from the Agence Nationale pour la Recherche, 
the program ANR-10-IDEX-0001-02 PSL and the state program Investissements d'avenir managed by Agence Nationale de la Recherche (Grant ANR-10-BINF-05Ò PherotaxisÓ).

\bibliography{bj_bibtex_template}

% closing statement, nothing below matters
\end{document}